\documentstyle[preprint,prb,aps]{revtex}
\begin{document}
\draft

\title{Short--range impurity in the vicinity of a  saddle point 
and  the levitation of the 2D delocalized states in a magnetic field.}
\author{A. Gramada and M. E. Raikh}
\address{Department of Physics, University of Utah, Salt Lake City, Utah  84112}

\maketitle

\begin{abstract}
The effect of a short--range impurity on the transmission through a 
saddle--point potential for an electron, moving in a strong magnetic field,
is studied. It is demonstrated that for a random position of an impurity
and random sign of its potential the impurity--induced mixing of the  Landau
 levels diminishes {\em  on average} the transmission coefficient. This 
results in an upward shift (levitation) of the energy position of the 
delocalized state in a smooth potential. The magnitude of the shift is
estimated. It increases with decreasing magnetic field $B$ as $B^{-4}$.   
\end{abstract}
\pacs{PACS Numbers: 73.20.Jc, 73.40.Hm}
\section{Introduction}
\label{sec:I}
The fate of the two--dimensional delocalized states (DS's) in a magnetic field
became recently the subject of intensive experimental
\cite {Gloz1,Gloz2,Gloz3,Kra,Fur} and theoretical \cite {SR,Liu,Kag} studies.
The goal of these studies is to trace the position of DS's as 
magnetic field  decreases. More than a decade ago  it was predicted
that with decreasing  magnetic field each DS  departs from the
center of the Landau level (LL) and floats up in energy (levitation scenario
 \cite {Khm,L}). 
This prediction was derived from  the non--interacting 
scaling theory of the quantum Hall effect \cite {LLP}. Since the position of
a DS determines the boundary between the insulating and the 
quantum Hall phases, such a floating implies that within some range of
concentrations the electron
 gas  undergoes an insulator--metal--insulator 
transition with decreasing magnetic field.
Later it was argued  that this behavior should
persist in the presence of interactions \cite {KLZ}. 
Experimental observation of the
insulator--metal--insulator transition was reported by several
groups \cite {J,JJ,W,H}.

It is obvious just from the electron--hole symmetry that the deviation of DS
from the center of Landau level is possible only due to
the disorder--induced LL  mixing.
Numerical simulations  \cite {Kag,Kun,An} indeed support
the levitation scenario. Analytical theory of the levitation was developed 
only for the region of magnetic fields where the departure of the DS 
from the center of LL is relatively small \cite{SR}. It was 
assumed  \cite {SR} that the random potential is smooth, so that the structure 
of electronic states is described by the network model of 
Chalker and Coddington \cite {CC}. In this model, delocalization results 
from the tunneling of an electron through the saddle points of a random 
potential which are connected by equipotential lines. It was demonstrated  
that the LL mixing changes {\em on average}  the transmittancy of 
a saddle point in such a way that it becomes smaller than  1/2  for  
the energy at the center of the LL. This means that to achieve the  
1/2  average transmittancy, the energy should be shifted upwards, which is 
equivalent to the levitation. However, in a smooth potential the LL
 mixing is generally weak since it is associated with the large 
momentum transfer. Short--range potential is much more effective in this 
respect. That is why in this paper we consider the situation when a small
portion of short--range impurities is present in a sample in 
addition to a smooth potential.
Within the Chalker--Coddington model the shift in position of the DS
results from the change in  the transmittancy of a saddle point
due to LL mixing. Then it is obvious that  impurities,
responsible for this shift, are those that fall in the vicinity of 
saddle points. We assume the sign of the impurity potential to be random and
show that the net effect of such impurities
is the reduction of the 
transmittancy at a given energy, and, hence, the levitation of DS's, 
even if the mixing of LL's  by a smooth potential is neglected.

The plan of the paper is the following. In Sec. \ref{sec:H} we consider
the motion of an electron in a strong magnetic field and a saddle--point
 potential with a short--range impurity located nearby. We derive
the  system of equations for the motion of the guiding center using
the procedure developed by Fertig and Halperin \cite{fer}.
In Sec. \ref{sec:R} we demonstrate that coupling of LL's  by an 
impurity leads to the renormalization of its effective strength.
In Sec. \ref{sec:T} a general expression for the transmission coefficient is
derived  and different limits are analyzed.
In Sec. V the impurity--induced change in the average transmittancy is studied.
In Sec. VI the magnitude of levitation of DS's, which results from the
reduction of the average transmittancy is, estimated. 

\section{Hamiltonian}
\label{sec:H}
The Hamiltonian we  consider has the form 
\begin{equation}
\label{1}
\hat{H}=\frac{1}{2m}\left(\frac{\hbar}{i}\nabla+\frac{e}{c}\bf A\right)^2+
V_{SP}(x,y)+V(x,y)\equiv \hat{H_0}+V(x,y),
\end{equation}
where
\begin{equation}
\label{2}
V_{SP}=-\frac{m\Omega^2}{2}(x^2-y^2)
\end{equation}
is a potential of the saddle point;
\begin{equation}
\label{3}
V(x,y)=V_0 \ F\left(\sqrt{(x-x_0)^2+(y-y_0)^2} \right) 
\end{equation}
is the short range potential of an impurity centered at the point $(x_0,y_0)$.
$V_0$ is chosen in such a way that $F(0)=1$.
We choose  a symmetric gauge, {\bf A}=$\frac{B}{2}(-y,x,0)$,
where $B$ is the magnetic field.
To calculate the transmission coefficient of the saddle point 
we  adopt the approach developed  by Fertig
and Halperin \cite{fer}. It was demonstrated \cite{fer}
that the Hamiltonian $\hat{H}_0$ can be
separated into two parts describing respectively
the motion of a guiding center and a cyclotron  motion.
The  unitary transformation which provides this separation is as follows:
\begin{eqnarray}
\label{4}
x &= &l \sqrt{2} \;(\cos\beta \; X+\sin\beta \; s),\; \; \; \; \;
\frac{\partial}{\partial x} = \frac{1}{l \sqrt{2}}\; \left (\cos\beta\; \frac{\partial}{\partial X}+\sin\beta \; \frac{\partial}{\partial s}\right ) \ ,\\
\label{5}
y &=&i l \sqrt{2}\;\left(\sin \beta \; \frac{\partial}{\partial X}-\cos \beta \;
\frac{\partial}{\partial s}\right), \; \;
\frac{\partial}{\partial y} = -\frac{i}{l \sqrt{2}}\;\left(\sin \beta \;X-
\cos \beta \;s \right),
\end{eqnarray}
where   $l=\sqrt{\hbar/m \omega_c}$   is the magnetic length
($\omega_c$ stands for the cyclotron frequency).
The new operators satisfy the commutation relations $\left[X,s \right]=
\left[\frac{\partial}{\partial X},\frac{\partial}{\partial s} \right]=0$,
$\left[\frac{\partial}{\partial X},X \right]=\left[\frac{\partial}{\partial s},
s \right]=1$.
The parameter $\beta$ is determined by the equation
\begin{equation}
\label{8}
\tan(2\beta)=-\frac{1}{2}\;\left(\frac{L}{l}\right)^4,
\end{equation}
where $L$ is the characteristic length defined as 
$L= \sqrt{ \hbar/m \Omega}$. The condition that the potential is smooth can be 
quantitatively expressed as  $\omega_c \gg \Omega$, which means that
$L \gg l$.
Then the solution of  Eq.(\ref{8}), can be written as
\begin{equation}
\label{10}
\beta = -\frac{\pi}{4}+\left(\frac{l}{L} \right)^4 + O\left( \left(l/L \right)^8 \right) \ .
\end{equation}
 As a result of the transformation (4)--(7), the Hamiltonian 
$\hat{H}_0$ takes the form
\begin{equation}
\label{11}
\hat {H_0}= E_1  \left(-\frac{\partial^2}{\partial X^2}-X^2 \right)+
\frac{1}{2} E_2  \left(-\frac{\partial^2}{\partial s^2}+s^2 \right),
\end{equation}
where the energies  $E_1, E_2$ in the limit $L \gg l$ are given by
\begin{equation}
\label{12}
E_1 =  \frac{\hbar \Omega^2}{2 \omega_c}, \  \ 
E_2  = \hbar \omega_c  \ .
\end{equation}

It is seen  that the variable $X$ describes the motion of the guiding center
whereas $s$ is responsible for the cyclotron motion.

We search for the solution of the Schr\"odinger equation, $\hat{H} \Psi=
E \Psi$, in the form
\begin{equation}
\label{15}
\Psi(x,y)=\sum \limits_{n} C_n(X) \psi_n(s),
\end{equation}
where $\psi_n$ are the  eigenfunctions of the harmonic oscillator
\begin{equation}
\label{16}
\left( -\frac{\partial^2}{\partial s^2}+s^2 \right) \psi_n(s)=
(2n+1) \psi_n(s) \ .
\end{equation}
The coefficients $C_n$ satisfy the system of equations 
\begin{eqnarray}
\label{17}
\frac{d^2C_n}{dX^2}&+&(X^2+\varepsilon_n) C_n= \nonumber \\
& & \nonumber \\ 
&=&  \frac{V_0}{E_1} \int \limits_{-\infty}^{\infty}ds \ 
\psi_n(s)  F\left(l (X-s)-x_0,-i l  \left( \frac{\partial}
{\partial X}+ \frac{\partial}{\partial s}\right)-y_0 \right) \Psi(X,s),
\end{eqnarray}
with $\varepsilon_n$ defined as
\begin{equation}
\label{18}
\varepsilon_n=\frac{E- \hbar \omega_c(n+1/2)}{E_1} \ .
\end{equation}

First, it can be readily seen that, since the function $F$ is nonzero only
when both arguments are smaller than $a$, where $a\ll l$ is the radius of
the impurity potential, one can take $\psi_n(s)$ out of the integral at 
point $s=X-\frac{x_0}{l}$. We also see that after switching to the $(X,s)$
variables the impurity potential became an operator. Thus, to find its
action on the function  $\Psi$ one should perform the Fourier 
transformation 
\begin{equation}
\label{18b}
\Psi(X,s)=\Psi \left(\frac{X+s}{2}+\frac{x_0}{2l},\frac{X+s}{2}-\frac{x_0}{2l}
\right)=\int \limits_{-\infty}^{\infty}dq \ e^{iq \frac{X+s}{2}}\Phi_q,
\end{equation}
where we again took into account that $X-s\approx \frac{x_0}{l}$. As a result
the differential  operator $\frac{\partial}{\partial X}+\frac{\partial}
{\partial s}$ in the argument of $F$ can be replaced by $iq$, so that 
the right--hand side takes the form 
\begin{equation}
\label{18c}
\frac{V_0}{E_1} \ \psi_n(X-x_0/l)\int \limits_{-\infty}^{\infty}ds
\int \limits_{-\infty}^{\infty}dq \ e^{iq \frac{X+s}{2}}\Phi_q \
F \left( l(X-s)-x_0,ql-y_0 \right) \ .
\end{equation}
Now the integration over $s$ and $q$ can be easily performed using the
short--range character of $F$. This picks the values $q=\frac{y_0}{l}$ 
and $s=X-\frac{x_0}{l}$ and generates the factor $\frac{\pi a^2}{l^2}$.
After that it is convenient to express the component 
$\Phi_{\frac{y_0}{l}}$ in terms of $\Psi$ using the inverse Fourier 
transformation. Finally we obtain
\begin{equation}
\label{19}
\frac{d^2C_n}{dX^2}+(X^2+\varepsilon_n) \ C_n = \tilde{V}_0 \ \lambda \ e^{iXY_0}
 \ \psi_n(X-X_0),
\end{equation}
where we have introduced the notations
\begin{equation}
\label{20}
\tilde{V}_0= \frac{1}{E_1} . \frac{V_0 a^2}{2 \ l^2}, \ \
X_0 = \frac{x_0}{l},\ \  Y_0=\frac{y_0}{l} \ . 
\end{equation}
The paremeter $\lambda$ is determined as
\begin{equation}
\label{21}
\lambda = \int \limits_{-\infty}^{\infty}ds \ e^{-isY_0} \ \Psi(s,s-X_0)
= \sum \limits_{n} \int \limits_{-\infty}^{\infty}ds\ e^{-isY_0} \ C_n(s)
 \psi_n(s-X_0) \ .
\end{equation}
In the second equality we have substituted $\Psi(s,s-X_0)$ using 
the representation (\ref{15}).
Eqs. (\ref{19}), (\ref{21}) form a closed system of equations describing
the motion of the guiding center.
In the absence of a short--range impurity, $V_0=0$, so that (\ref{19})
reduces to the system of identical Schr\"odinger equations for a particle
in an inverted parabolic potential. The transmission coefficient for
this potential is well known: 
\begin{equation}
\label{22}
T_0(\varepsilon_n)=\frac{1}{1+e^{-\pi \varepsilon_n}} \ \ \ ,
\end{equation}
and is independent of the number of LL.

\section{Renormalization of the Scattering Strength}
\label{sec:R}
Since we are interested in scattering of an electron with energy close 
to the center of some LL, say $n_0$, by the saddle point potential, we 
can assume $\varepsilon_m \approx (n_0-m)\frac{\hbar \omega_c}{E_1}$ for
any $m \ne n_0$. The condition $\omega_c \gg \Omega$ guarantees that 
$\hbar\omega_c \gg E_1$. This means that each of Eqs. (\ref{19})
 with $m \ne n_0$ contains a big parameter $\varepsilon_m \gg 1$. Then for
$m \ne n_0$ one can neglect the first two terms in (\ref{19}) and get the
solution
\begin{equation}
\label{renorm}
C_n=\lambda\frac{\tilde{V}_0}{\varepsilon_n} e^{iXY_0}
 \ \psi_n(X-X_0).
\end{equation}
Substituting this solution into Eq. (\ref{21}), enables us to
 express the constant
$\lambda$ in terms of only one unknown function $C_{n_0}(X)$
\begin{equation}
\label{lam}
\lambda = \frac{ \int \limits_{-\infty}^{\infty}ds \  e^{-isY_0} \ C_{n_0}(s)
\ \psi_{n_0}(s-X_0)}{1-\tilde V_0 \sum \limits_{m \not= n_0}\frac{1}{\varepsilon_m}}.
\end{equation}
Once $\lambda$ is determined, one can write a closed equation 
for $C_{n_0}$
\begin{equation}
\label{28}
\frac{d^2C_{n_0}}{dX^2}+(X^2+\varepsilon_{n_0}) \ C_{n_0} = \tilde{V}_0^R  \
e^{iXY_0} \ \psi_{n_0}(X-X_0) \int \limits_{-\infty}^{\infty}ds \  e^{-isY_0} \ C_{n_0}(s) \ \psi_{n_0}(s-X_0) \ ,
\end{equation}
where  $\tilde{V}_0^R$ is defined as
\begin{equation}
\label{30}
\tilde{V}_0^R=\frac{\tilde{V}_0}{1+\tilde{V}_0K}
\end{equation}
with $K$ given by
\begin{equation}
\label{K}
K=-\sum \limits_{m \not= n_0}\frac{1}{\varepsilon_m}
\end{equation}
We see that the parameter $K$ describes the effect of all other LL's
on the motion of an electron on the level $n_0$. It is apparent 
from (\ref{30}) that
this effect reduces to the renormalization of the scattering strength
$\tilde{V}_0$. Important is that, since the  impurity is 
short--range,  it causes a strong local mixing of LL's so that many 
levels with $m \ne n_0$ contribute to the renormalization constant $K$. 
Indeed, the terms in the sum (\ref{K}) fall off with the number of LL as
$\frac{1}{m}$.
In the case $n_0=0$ all the terms in the sum (\ref{K}) have the same sign,
 so, strictly speaking, the sum diverges logarithmicaly.
The cutoff value of $m$ is determined by the following condition.
For large enough numbers of LL's $\psi_m(s)$ is a rapidly oscillating
function, the typical period of oscillations being $\sim 1/ \sqrt{m}$.
Taking $\psi_m(s)$ out of integral in (\ref{17}) is justified only
if this period is much larger than the dimensionless size of the 
impurity potential $a/l$. This leads to the cutoff value 
$m_{max} \sim  l^2/a^2$. Thus we obtain 
\begin{equation}
\label{KAA}
K \approx  \frac{E_1}{\hbar \omega_c} \ln m_{max} = 
2 \frac{E_1}{\hbar \omega_c} \ln \frac{l}{a} \  .
\end{equation}
For $n_0 \not = 0$ the sum in (\ref{K}) contains both positive and 
negative contributions. Positive terms are those with $m < n_0$, 
while the terms with $m > n_0$ are negative. For large enough $n_0$ 
(but much smaller than $m_{max}$) we have 
\begin{equation}
\label{KAAA}
K \approx  \frac{E_1}{\hbar \omega_c} \ln \frac{m_{max}}{n_0} = 
\frac{E_1}{\hbar \omega_c} \ln\frac{l^2}{a^2 n_0} \ .
\end{equation}
Thus the  magnitude  of the renormalization constant $K$ 
decreases with the number of LL.

\section{Transmission coefficient}
\label{sec:T}
In this section we will drop for simplicity
the LL index of  the transmission coefficient.
Without an impurity, the solutions of Eq. (\ref{28}) 
are the  parabolic cylinder functions \cite{A.E}
$D_{\nu}(X \sqrt{2} e^{\frac{i\pi}{4}})$ and 
$D_{\nu}(-X \sqrt{2} e^{\frac{i\pi}{4}})$,
where $\nu = - \frac{i \varepsilon}{2} - \frac{1}{2}$.
Using the asymptotic form of $D_{\nu}$
\begin{eqnarray}
\label{AS1}
D_{\nu}(X) &\sim&  X^{\nu}e^{-\frac{1}{4} X^2}, \ \ \  X \rightarrow \infty, 
 \ \ \ \left( - \frac{3 \pi}{4}< argX < \frac{3 \pi}{4} \right) \\
\label{AS2}
D_{\nu}(X) &\sim& X^{\nu}e^{-\frac{1}{4} X^2}-
\frac {(2 \pi)^{1/2}}{\Gamma (- \nu)}e^{- i \pi \nu }X^{- \nu -1}
e^{\frac{1}{4} X^2}, \ \ \  X \rightarrow - \infty, \ \ \ 
\left(- \frac{5 \pi}{4}< argX < -\frac{\pi}{4}\right), 
\end{eqnarray}
one can  see that it is the function 
$D_{\nu}(X \sqrt{2} e^{\frac{i\pi}{4}})$ that has the  ''right''
behavior at $ X \rightarrow \pm \infty $ ( no incoming 
wave at $ X \rightarrow + \infty$). The amplitude transmission 
coefficient, $t$, emergies from the comparison of the amplitudes 
of the transmitted and the incoming waves in (\ref{AS1}),
(\ref{AS2}) 
\begin{equation}
\label{TA}
t=\frac{\left( \sqrt{2} e^{i \frac{\pi}{4}} \right)^{2 \nu +1}}
{(2 \pi)^{1/2}} \  \Gamma (- \nu)=
\frac{e^{\frac{ \pi \varepsilon}{4}- 
\frac{i \varepsilon}{2} \ln 2}}{(2 \pi)^{1/2}} \ 
\Gamma \left(  \frac{i \varepsilon}{2}+ \frac{1}{2} \right) 
\end{equation}
It is easy to see that $ |t|^2 = T_0(\varepsilon) $, 
where  $T_0(\varepsilon)$ is given by (\ref{22}).
With the right--hand side in (\ref{28}) present, it is convenient 
to search for a  solution of Eq. (\ref{28}) in the form
 of a linear combination 
of the functions $D_{\nu}$
\begin{equation}
\label{43}
C_n(X)= D_{\nu}\left(X \sqrt{2}e^{i \frac{\pi}{4}} \right)+
\int \limits_{- \infty}^{\infty}d \varepsilon^{\prime} \left[
a^+(\varepsilon^{\prime})\
D_{\nu^{\prime}} \left(X \sqrt{2}e^{i \frac{\pi}{4}}\right)+
a^-(\varepsilon^{\prime})\ D_{\nu^{\prime}}
\left(-X \sqrt{2}e^{i \frac{\pi}{4}} \right) \right] \ .
\end{equation}

The expressions for the coefficients $a^{\pm}$ are obtained 
by multiplying Eq.  (\ref{28}) by 
$D_{\nu}^*(\pm X \sqrt{2}e^{i \frac{\pi}{4}})$
and integrating over $X_0$. One has
\begin{equation}
\label{44}
a^{\pm}(\varepsilon^{\prime})= \frac{\tilde{V}_0^R}{\sqrt{2}} \ 
\frac{J \ (I^{\pm}(\varepsilon^{\prime}))^*}
{\alpha (\varepsilon^{\prime}) (\varepsilon- \varepsilon^{\prime})},\\
\end{equation}
where $I^{\pm}$ and $J$ are defined as
\begin{eqnarray}
\label{Iint}
I^{\pm}(\varepsilon)&=&\int \limits_{- \infty}^{\infty}dX \ e^{-iXY_0} 
 \psi_n(X-X_0) \
D_{\nu}\left(\pm X \sqrt{2}e^{i \pi /4}\right ), \\
\label{Jint}
J &=& \int \limits_{- \infty}^{\infty}ds \ e^{-isY_0}  
C_n(s)  \psi_n(s-X_0),
\end{eqnarray}
and the function $\alpha (\varepsilon)$ is determined by the normalization 
condition
\begin{equation}
\label{N}
\int \limits_{- \infty}^{\infty}d \xi\  D_{\nu}^* 
\left(\pm \xi \sqrt{2}e^{i \pi /4}\right ) 
D_{\nu^{\prime}} 
\left(\pm \xi \sqrt{2}e^{i \pi /4}\right )=2 \alpha (\varepsilon) \
\delta (\varepsilon-\varepsilon^{\prime}) .
\end{equation}
Using the asymptotics (\ref{AS1}), (\ref{AS2}) it can be shown that
\begin{equation}
\label{NC}
\alpha(\varepsilon)= 2 \pi \left(e^{\frac{\pi \varepsilon}{2}}+
e^{-\frac{\pi \varepsilon}{2}} \right)  
e^{-\frac{\pi \varepsilon}{4}}.
\end{equation}
Upon substituting (\ref{44}) into (\ref{43}), we obtain
\begin{eqnarray}
\label{49}
C_n(X)&=& D_{\nu}\left(X \sqrt{2}e^{i \pi /4}\right )+ \nonumber \\
& &J \frac{\tilde{V}_0^R}{\sqrt{2}}\int \limits_{- \infty}^{\infty} 
\frac{d\varepsilon^{\prime}\alpha^{-1} (\varepsilon^{\prime})}
{(\varepsilon-\varepsilon^{\prime})} 
\left[ (I^{+}(\varepsilon^{\prime}))^* 
D_{\nu^{\prime}} \left(X \sqrt{2}e^{i \frac{\pi}{4}}\right) 
+ (I^{-}(\varepsilon^{\prime} ))^* D_{\nu^{\prime}}
\left(-X \sqrt{2}e^{i \frac{\pi}{4}}\right) \right].
\end{eqnarray}
For $X \rightarrow \pm \infty$ only the poles contribute 
to the integrals in Eq. (\ref{49}). In the limit 
$X \rightarrow \infty $, 
$D_{\nu^{\prime}} (X \sqrt{2}e^{i \frac{\pi}{4}}) $ 
is just a transmitted   wave, whereas 
$D_{\nu^{\prime}} (-X \sqrt{2}e^{i \frac{\pi}{4}})$ represents
the combinations of waves traveling in both directions.
Collecting the components corresponding to the transmission, 
we obtain the following general expression for the 
transmission coefficient
\begin{equation}
\label{51}
T(\varepsilon)=T_0(\varepsilon) 
\left| 1+i \pi \ \sqrt{2} \ J \ \tilde{V}_0^R \  
\alpha^{-1} \left(\varepsilon\right) 
\left[ (I^{+}(\varepsilon))^*+i e^{- \frac{\pi \varepsilon}{2}}
(I^{-}(\varepsilon))^* \right] \right|^2 .
\end{equation}
To find the constant $J$ one should multiply (\ref{49}) by
$e^{-iXY_0} \psi_n(X-X_0)$ and integrate. This will generate $J$ in
the left-hand side; the solution of the resulting equation for $J$ has the form
\begin{equation}
\label{47}
J= \frac{I^+(\varepsilon)}{1-
\frac{\tilde{V}_0^R}{\sqrt{2}}\int \limits_{- \infty}^{\infty} 
\frac{d\varepsilon^{\prime}} {(\varepsilon-\varepsilon^{\prime})} \ 
\alpha^{-1} (\varepsilon^{\prime}) 
\left[\left|I^+(\varepsilon^{\prime}) \right|^2 +
\left|I^- (\varepsilon^{\prime} ) \right|^2 \right]} .
\end{equation}
With $J$ given by (\ref{47}), the final result for the transmission
coefficient reads
\begin{equation}
\label{TG}
T(\varepsilon)=T_0(\varepsilon) \left| 1+
\frac{i \pi \sqrt{2} \  \tilde{V}_0^R  \ 
\alpha^{-1} (\varepsilon) 
\left[ \left|I^+(\varepsilon)\right|^2+
i e^{- \frac{\pi \varepsilon}{2}}
I^{+}(\varepsilon)(I^{-}(\varepsilon))^* \right]}
{1-\frac{\tilde{V}_0^R}{\sqrt{2}}\int \limits_{- \infty}^{\infty} 
\frac{d\varepsilon^{\prime}} {(\varepsilon-\varepsilon^{\prime})} \ 
\alpha^{-1} (\varepsilon^{\prime}) 
\left[\left|I^+(\varepsilon^{\prime}) \right|^2 +
\left|I^- (\varepsilon^{\prime} ) \right|^2 \right]}  \right|^2 .
\end{equation}

The denominator of (\ref{TG}) can be presented as
$\left(1-\tilde{V}_0^R \Sigma(\varepsilon) \right)-
i \tilde{V}_0^R \Theta(\varepsilon)$, 
where the functions $\Sigma$ and $\Theta$ are defined as
\begin{eqnarray}
\label{SIGMAF}
\Sigma(\varepsilon)&=&\frac{1}{\sqrt{2}} 
P\int \limits_{- \infty}^{\infty} 
\frac{d\varepsilon^{\prime}} {(\varepsilon-\varepsilon^{\prime})} \ 
\alpha^{-1} (\varepsilon^{\prime}) 
\left[\left|I^+(\varepsilon^{\prime}) \right|^2 +
\left|I^- (\varepsilon^{\prime} ) \right|^2 \right] \ , \\
\label{TETAF}
\Theta( \varepsilon)&=& \frac{\pi}{\sqrt{2}} \ 
\alpha^{-1} (\varepsilon) 
\left[\left|I^+(\varepsilon) \right|^2 +
\left|I^- (\varepsilon) \right|^2 \right]\ , 
\end{eqnarray}
where the symbol $P$ stands for principal part.
The meaning of the functions $\Sigma$ and $\Theta$ can be understood 
in the following way. 
It is known that without a smooth potential, a short--range 
impurity pulls out a state  from the center of a degenerate 
LL and forms a bound state \cite{A,B,GM,Gr,Az}.
 In the presence of a saddle--point 
potential the energy position of  the bound state is determined 
by the condition: \hspace{1mm} $1-\tilde{V}_0^R \Sigma(\varepsilon)=0$. On the 
other hand, since the potential (\ref{2}) ``bends'' the LL, 
the bound state becomes degenerate with the continuum of states at the 
 LL  and, thus, 
acquires a finite width which is described by the function 
$\Theta(\varepsilon)$. It is seen that $\Theta(\varepsilon)$ 
represents the sum of two terms, which correspond to the widths 
associated with the outcome to the left and to the right, 
respectively.
When the impurity is located exactly at the saddle point
($X_0=Y_0=0$) we have $I^+(\varepsilon)=I^-(\varepsilon)$, so that
(\ref{TG}) reduces to
\begin{equation}
\label{T0}
T(\varepsilon)=T_0(\varepsilon) \frac{\left[(1-\tilde{V}_0^R \ \Sigma(\varepsilon))
-\tilde{V}_0^R \ \Theta(\varepsilon) \ e^{- \frac{\pi \varepsilon}{2}}  
\right]^2 }{(1-\tilde{V}_0^R \ \Sigma(\varepsilon))^2+
(\tilde{V}_0^R \ \Theta(\varepsilon))^2}.
\end{equation}
For this particular location of the impurity, the functions $\Sigma$ 
and $\Theta$ are, respectively, odd and even functions of energy.
We have evaluated these functions for the lowest LL. Note first that for 
$n=0$ and arbitrary $X_0,Y_0$  the integrals $I^+$ and $I^-$ can
 be expressed in terms of the 
parabolic cylinder functions if one substitutes into (\ref{Iint}) the
integral representation \cite{A.E} of $D_{\nu}(X)$. Then one gets
\begin{eqnarray}
\label{ID}
I^+(\varepsilon,X_0,Y_0)&=&(2\pi)^{1/4}e^{-\frac{1}{4}(X_0^2 + Y_0^2 + 2iX_0Y_0
+\frac{\pi\varepsilon}{2})}D_{\nu}\left(e^{\frac{\pi i}{4}}(X_0-iY_0)\right); \nonumber
\\
& &I^-(\varepsilon,X_0,Y_0)=I^+(\varepsilon,-X_0,-Y_0).
\end{eqnarray}  
Since $D_{\nu}(0)=2^{-(\nu/2 +1)}\Gamma(-\frac{\nu}{2})/\Gamma(-\nu)$, where
$\Gamma(z)$ is the $\Gamma$--function, we get the following expression for
$I^+=I^-$ at $X_0=Y_0=0$
\begin{equation}
\label{I0}
I^+(\varepsilon)=I^-(\varepsilon)= \pi^{\frac{1}{4}} 
2^{\frac{i \varepsilon}{4}-\frac{1}{2}}
e^{-\frac{\pi \varepsilon}{8}} \ 
\frac{\Gamma \left( \frac{i \varepsilon}{4}+ \frac{1}{4} \right) }
{\Gamma \left( \frac{i \varepsilon}{2}+ \frac{1}{2} \right)}
\end{equation}
Then the functions $\Theta$ and $\Sigma$ take the form
\begin{equation}
\label{TETA0}
\Theta(\varepsilon)= \frac{1}{4\sqrt{2 \pi}}
\left| \Gamma \left( \frac{i \varepsilon}{4}+ \frac{1}{4} \right) \right|^2 \ .
\end{equation}
\begin{equation}
\label{SIG}
\Sigma(\varepsilon)=\frac{1}{\pi}\int \limits_{- \infty}^{\infty}d\varepsilon^{\prime}\frac{\Theta(\varepsilon^{\prime})}{\varepsilon - \varepsilon^{\prime}} =
\frac{1}{2\sqrt2}\int \limits_{0}^{\infty}dv \frac {e^{-\frac{v}{4}}\sin(\frac{\varepsilon v}{4})}
{\sqrt{1 + e^{-v}}}.
\end{equation}
$\Theta(\varepsilon)$ is shown in Fig. 1 together with
 the function $\Sigma$ calculated
numerically. The asymptotic behavior of $\Sigma(\varepsilon)$ is as follows:
$\Sigma(\varepsilon) = 1.4\varepsilon$, for $|\varepsilon| \ll 1$; $\Sigma(\varepsilon)= 
1/\varepsilon$, for $|\varepsilon| \gg 1$. The energy dependence of the transmission
coefficient for different amplitudes of the impurity potential, $\tilde V_0^R$,
 is shown in Fig. 2.
At  zero energy   $\Sigma(\varepsilon)$ vanishes, so that 
(\ref{T0}) simplifies to
\begin{equation}
\label{T00}
T(\varepsilon=0)=\frac{1}{2} \  \frac{\left(1-
\frac{\tilde{V}_0^R}{4 \sqrt{2 \pi}} 
\left|\Gamma (\frac{1}{4})\right|^2 \right)^2} 
{1+ \frac{\left( \tilde{V}_0^R \right)^2} {32 \pi} 
\left|\Gamma (\frac{1}{4})\right|^4} \ .
\end{equation}
Fig. 3 shows the transmission coefficient at $\varepsilon=0$ as a function of
$\tilde V_0^R$.

\section{Reduction of the average transmittancy}
\label{S}

Without an impurity, $T_0(\varepsilon)$ obeys the relation:
$T_0(\varepsilon)+T_0(-\varepsilon)=1$ which is the manifestation 
of the symmetry of the saddle point (transmission of an 
electron with energy $\varepsilon$ can be viewed as a reflection for an electron 
with energy $-\varepsilon$). Obviously, when the impurity is
randomly positioned around the saddle point 
and the sign of $\tilde{V}_0^R$ is random, 
this symmetry relation should hold {\it on average}.
Indeed, it can be demonstrated that for an arbitrary position of an impurity,
$(X_0, Y_0)$, the transmission coefficient satisfies the relation
\begin{equation}
\label{GS}
T(\varepsilon,\tilde{V}_0^R, X_0,Y_0)+
T(-\varepsilon ,-\tilde{V}_0^R, -Y_0,-X_0)=1
\end{equation}
To prove (\ref{GS}), one should rewrite (\ref{TG})
in the form
\begin{equation}
\label{GTC}
T(\varepsilon)=T_0(\varepsilon) \frac{\left|(1-
\tilde{V}_0^R \ \Sigma (\varepsilon))+i\tilde{V}_0^R \ \Pi(\varepsilon)
-\tilde{V}_0^R \ \Lambda(\varepsilon) \ e^{- \frac{\pi \varepsilon}{2}}  
\right|^2 }{(1-\tilde{V}_0^R \ \Sigma(\varepsilon))^2+
(\tilde{V}_0^R \ \Theta(\varepsilon))^2} \ ,
\end{equation}
where the functions $\Pi$ and $\Lambda$ are defined as
\begin{eqnarray}
\label{A}
\Pi(\varepsilon)&=& \frac{\pi}{\sqrt{2}} \ \alpha^{-1} (\varepsilon) \ 
\left[ \left|I^+ ( \varepsilon ) \right|^2 -
\left|I^- (\varepsilon) \right|^2 \right], \\
\label{B}
\Lambda(\varepsilon)&=&\pi \sqrt{2} \ \alpha^{-1} (\varepsilon)  
\ I^+(\varepsilon)(I^- ( \varepsilon )).^*
\end{eqnarray}
The function $\Pi$ is real and turns to zero for an impurity located at  
$X_0=Y_0=0$; $\Lambda(\varepsilon)$ is, generally speaking, a complex function.
It is obvious that $\Pi^2+|\Lambda|^2= \Theta ^2$.
One can verify  that under  the transformation 
$(\varepsilon, X_0,Y_0) \rightarrow
(-\varepsilon , -Y_0,-X_0)$ the integrals $I^+$  and $I^-$ with an accuracy of
phase factors are transformed into $ e^{\frac{\pi \varepsilon}{4}}(I^-)^*$ 
and $ e^{\frac{\pi \varepsilon}{4}}(I^+)^*$ respectively; with the same accuracy
the conjugated values $(I^+)^*$ and $(I^-)^*$ are transformed into 
$ e^{-\frac{\pi \varepsilon}{4}}I^-$ 
and $ e^{-\frac{\pi \varepsilon}{4}}I^+$.
Then the rules of transformation for the functions $\Sigma,\Theta, \Pi$, and
$\Lambda$ are as follows:
$\Pi\rightarrow -\Pi , \  \Sigma \rightarrow  - \Sigma, \  
\Theta \rightarrow \Theta, 
 \ \Lambda \rightarrow \Lambda$. Using these properties, the relation (\ref{GTC}) can be easily proved. The meaning of this relation is that the transmittancy
of the saddle point, averaged over the amplitude $\tilde V_0^R$, position
$(X_0, Y_0)$, and energy $\varepsilon$ with  any {\em symmetric}
 distribution function,
is equal to $1/2$. The averaging over $\varepsilon$ has the meaning of averaging
over the background value of the saddle point potential, $V_{SP}$, (this value was set zero in (\ref{2})).

Our main observation is that for a symmetric distribution of the ``bare'' 
amplitudes $\tilde V_0$, the distribution of renormalized amplitudes, $\tilde
V_0^R$, which are  defined by Eq.~(\ref{30}), is {\em asymmetric}.
 The origin of
 the asymmetry is the LL mixing. Indeed,
 denote with $\phi(\tilde V_0)$ the distribution function of  $\tilde V_0$ \
 (we assume that $\phi(\tilde V_0)= \phi(-\tilde V_0)$).
Then using (\ref{30}) one can easily find the distribution function
  of $\tilde V_0^R$
\begin{equation}
\label{Dist}
\tilde\phi(\tilde V_0^R)=\frac{1}{(1-K\tilde V_0^R)^2}
\phi\left(\frac{\tilde V_0^R}{1-K\tilde V_0^R}\right).
\end{equation}
The degree of asymmetry of $\tilde \phi$, due to the finite renormalization constant $K$, is determined by the product $K\overline{V}$, where $\overline{V}$ is the width of
the distribution function $\phi$. Consider, for example, the Lorentzian
distribution: $\phi=\pi^{-1}\overline{V}/(\tilde V_0^2 + \overline{V}^2)$. 
Then $\tilde \phi$ takes the form
\begin{equation}
\label{phi}
\tilde\phi(\tilde V_0^R)=\frac{\overline{V}(1+K^2\overline{V}^2)}{\pi\left[
\left((1+K^2\overline{V}^2)\tilde V_0^R - K\overline{V}^2\right)^2 + \overline{V}^2\right]}.
\end{equation}
We see that $\tilde\phi$ is also a Lorentzian, but it is centered at
$\tilde V_0^R = K\overline{V}^2/(1+K^2\overline{V}^2)$. For
 $K\overline{V} \ll 1$ the shift of the center, $K\overline{V}^2$, is relatively small as compared to the width
$\overline{V}$. However for $K\overline{V} \gg 1$ we get a narrow peak at
$\tilde V_0^R =K^{-1}$ with the width $(K^2\overline{V})^{-1} \ll K^{-1}$. This
means that all the impurities, which were with equal probability repulsive and
attractive before the renormalization, become effectively repulsive after the renormalization.

The asymmetry in $\tilde V_0^R$ makes the average transmittancy of the saddle 
point smaller than $1/2$. To demonstrate it, it is convenient to study the 
combination $T(\varepsilon,X_0,Y_0)+
T(-\varepsilon,-Y_0,-X_0)$ for the same $\tilde V_0^R$. Moreover, since $K \ll 1$,
the renormalized impurity strength $\tilde V_0^R = K^{-1}$ is much larger than
$1$, so that in calculating this combination one can expand (\ref{GTC}) with
respect to the small parameter $1/\tilde V_0^R$. Using the transformation
 rules for  the functions $\Sigma,\ \Theta,\ \Pi$ and $\Lambda$, established
above, we obtain
\begin{equation}
\label{exp}
T(\varepsilon,X_0,Y_0)+T(-\varepsilon,-Y_0,-X_0) = 
 1-\frac{2}{\tilde V_0^R (\Sigma^2 +
\Theta^2)^2\cosh \frac{\pi \varepsilon}{4}}\Delta(\varepsilon,X_0,Y_0),
\end{equation}
where $\Delta(\varepsilon,X_0,Y_0)$ is defined as
\begin{equation}
\label{delta}
\Delta(\varepsilon,X_0,Y_0)=  (\Theta^2-\Sigma^2)
\mbox{{\em Re}}\Lambda + 2\Pi\Sigma\mbox{{\em Im}}\Lambda +
 2|\Lambda|^2\Sigma\sinh
 \frac{\pi \varepsilon}{2}.
\end{equation}
We see that it is the sign of the  parameter $\Delta$ that determines the  
sign of deviation of the average transmittancy from $1/2$. 

Consider first the simplest case $X_0=Y_0=0$. Then we have $\Pi=0$,
 {\em Im}$\Lambda=0$ and {\em Re}$\Lambda=\Theta$. Hence, the parameter
$\Delta$ reduces to
\begin{equation}
\label{Delt}
\Delta(\varepsilon,0,0)=\Theta\left[(\Theta^2-\Sigma^2)+2\Theta\Sigma\sinh
\frac{\pi \varepsilon}{2}\right].
\end{equation}
It follows from (\ref{Delt}) that $\Delta$ is positive if the condition 
$\Theta(\varepsilon)>|\Sigma(\varepsilon)|e^{-\frac{\pi|\varepsilon|}{2}}$ is met.
Fig. 1 illustrates that this  indeed is the case for any energy $\varepsilon$.
 
As the displacement $(X_0, Y_0)$  increases, the parameter
$\Delta$ falls off due to the general decay of the integrals $I^+, I^-$. 
However,  $\Delta$ remains positive on average. The easiest way to see it
is to consider the limit of large displacements: $(X_0^2 + Y_0^2) \gg 1$.
In this limit $I^+$ and $I^-$ acquire big phases. This results in a big phase
of the function $\Lambda\propto I^+(I^-)^*$. Consequently, {\em Re}$\Lambda$
and {\em Im}$\Lambda$ oscillate rapidly with the change of the impurity
position (roughly as $\sin(\frac{X_0^2-Y_0^2}{2})$  and 
$\cos(\frac{X_0^2-Y_0^2}{2})$). Thus the contributions of the first two terms
in (\ref{delta}) average out. When considering the third term, we note
that $|\Lambda|^2$ depends only
on the magnitude of the displacement (it can be shown that $|\Lambda| \propto
(X_0^2 +Y_0^2)^{-\frac{1}{2}}e^{-\frac{X_0^2 + Y_0^2}{2}}$), i.e. $|\Lambda|^2$
is constant if $X_0^2 + Y_0^2$ is constant. The form of the function $\Sigma$
depends on the angular position of the impurity when $X_0^2 + Y_0^2$ is fixed,
but, averaged over the angular position, $\Sigma$ is an odd function of energy.
One can check formally that $\Sigma(\varepsilon,X_0,Y_0)+\Sigma(\varepsilon,Y_0,X_0)$
changes sign when $\varepsilon$ changes sign. Therefore, the product 
$\Sigma\sinh\frac{\pi\varepsilon}{2}$ is positive on average. Thus we conclude
that average $\Delta$ is positive both for small and large displacements, and,
consequently, the average transmittancy of the saddle point is diminished due
to the impurity--induced LL mixing.

\section{Conclusion}
\label{con}

In order to compensate for the reduction of the average transmittancy, the 
energy position of the delocalized state in a smooth potential shifts up
with respect to the center of LL. Let us estimate this shift (levitation).
Denote with $n$ the concentration of the impurities. Clearly, to produce a
significant effect on the transmittancy, the impurity should be located within
the interval of the order of magnetic length from the center of the saddle
point ($X_0\sim Y_0\sim 1$). The probability to find such an impurity is $\sim
 nl^2$. For  $X_0\sim Y_0\sim 1$ we have $\Sigma\sim\Theta\sim\Delta\sim 1$ 
in Eq. (\ref{exp}). Thus, the magnitude of the reduction of the average
transmittancy is of the order of $(\tilde V_0^R)^{-1}$.  On the other hand,
we have established that if an  impurity is strong enough, the LL mixing 
renormalizes its amplitude to $V_0^R =K^{-1}$. As a result,
in the presence of impurities, the delocalized 
state corresponds to such an energy for which, in the absence of impurities,
the average transmittancy of the saddle point exceeds $1/2$ by $\sim nl^2K$.
Since without an impurity  the transmission coefficient $T_0$ has the 
energy scale $E_1$, determined by  Eq. (\ref{12}), 
we get the following estimate for the magnitude of the levitation $\delta E$
\begin{equation}
\label{levit}
\frac{\delta E}{E_1}\sim nl^2K.
\end{equation}
It is instructive to compare the levitation with the spacing,
$\hbar\omega_c$, between the LL's. One gets
\begin{equation}
\label{relat}
\frac{\delta E}{\hbar\omega_c}\sim nl^2K\frac{E_1}{\hbar\omega_c}
\sim nl^2\left(\frac{E_1}{\hbar\omega_c}\right)^2\ln\frac{l}{a}
\sim nl^2\left(\frac{\Omega}{\omega_c}\right)^4 \ln\frac{l}{a},
\end{equation}
where we have used the expression (\ref{KAA}) for $K$. In our consideration
we have assumed that there is only a single impurity near the the saddle 
point, which implies that $nl^2 \ll 1$. We have also assumed that $\Omega
\ll \omega_c$. Then the above estimate shows that under the assumptions
adopted the relative levitation is small. It also shows that the magnetic 
field dependence of the levitation is $\delta E \propto B^{-4}$.

In numerical simulations \cite{Kag} the authors restricted 
the  study  to
the two lowest LL's. They found that, due to the LL mixing, the
 lower DS shifts up
whereas the upper DS shifts down in energy (``attraction'' of DS's).
 Our result is consistent with
this observation. Indeed, if only two LL's are considered, the renormalization
constant, $K$, would be  positive for the lower level and negative for the
upper level.

Note in conclusion, that our main result -- the reduction of the 
transmittancy due to the LL mixing, was derived in the limit of
the strong  renormalization of the impurity potential. The criterion      
for that, formulated in the previous Section, is $K\overline{V} > 1$,
$\overline{V}$ being the typical value of the dimensionless potential
$\tilde V_0$. Using Eqs. (\ref{KAA}) and (\ref{20}) this criterion can be
rewritten in  terms of the ``bare'' amplitude and size of the impurity
potential as $V_0a^2\ln\frac{l}{a} > \hbar\omega_cl^2\sim \hbar^2/m$.
On the other hand, in our approach we have treated the impurity as 
point--like \cite{GrAz} and, thus, neglected the change of the
electron wave function within the radius $a$.
This is  justified if the condition $V_0<\hbar^2/ma^2$ is met.
The second  condition seems to restict the validity of the theory to
the region $1>V_0/(\frac{\hbar^2}{ma^2})>1/\ln\frac{l}{a}$. It appears,
however, that the condition $V_0/(\frac{\hbar^2}{ma^2})< 1$ is not relevant.
We address this question in the  Appendix and  show that 
as soon as  $V_0>\hbar^2/ma^2\ln\frac{l}{a}$, it renormalizes to
$\hbar^2/ma^2\ln\frac{l}{a}$ \ (which is equivalent to $\tilde V_0^R=K^{-1}$), 
as we have assumed.

\appendix
\section*{}

Since the presence of the saddle--point potential is not important for the
renormalization procedure, we will trace the renormalization for the case
when $V_{SP}$ is absent. In this case the effect of an impurity is just the 
formation of the bound states which split off the LL's. In the symmetric
gauge with an impurity at the origin, the Schr\"odinger equation allows
 the separation of variables. The
 radial wave function, $R(\rho)$, for a zero orbital moment, satisfies
 the equation  
\begin{equation}\label{a1}
\frac{d^2R}{d\rho^2} + \frac{1}{\rho}\frac{dR}{d\rho}+\left[\frac{2m}{\hbar^2}
(E-V_0F(\rho))-\frac{m^2\omega_c^2}{4\hbar^2}\rho^2\right]R=0.
\end{equation}
We will analyze the solutions of (\ref{a1}), corresponding to high LL's,
for which the calculations can be performed semiclassically without invoking
the hypergeometric function. Then the  Bohr-Sommerfeld condition for the
energy levels, $E_p$, reads
\begin{equation}\label{a2}
E_p=\hbar\omega_c\left(p+\frac{1}{2}-\frac{\varphi_p}{\pi}\right),
\end{equation}
where $\varphi_p$ is an additional phase shift 
acquired at the  origin ($\varphi_p=0$
 for $V_0=0$). This shift should be 
found by  matching the solutions at $\rho < a$ and at $\rho > a$.      
Important is that for large $p$ the magnetic potential, $m\omega_c^2\rho^2/8$,
in (\ref{a1}) comes into play only at large $\rho\sim lp^{1/2}$. For smaller
$\rho$ one can neglect the magnetic potential. Then the solution of (\ref{a1})
(which is finite at $\rho=0$) can be written as
\begin{eqnarray}\label{a3} 
R(\rho)={\cal J}_0(q\rho),\ \ \ \rho<a \nonumber \\
R(\rho)=\nu_1{\cal J}_0(k\rho)+\nu_2{\cal N}_0(k\rho), \ \ \rho>a,
\end{eqnarray}
where ${\cal J}_0$ and ${\cal N}_0$ are, respectively, the Bessel and the 
Neumann functions of order zero; $k=(2mE_p)^{1/2}/\hbar$, and
 $q=(2m(E_p-V_0))^{1/2}/\hbar$. The constants $\nu_1, \nu_2$
determine the phase shift $\varphi_p$. Indeed, at $k\rho \gg 1$ the functions
${\cal J}_0$ and ${\cal N}_0$ oscillate as $\sin(k\rho-\frac{\pi}{4})$ and
  $\cos(k\rho-\frac{\pi}{4})$, so that
$\varphi_p = \arctan(\nu_2/\nu_1)$. The continuity conditions for $R(\rho)$ 
and $dR/d\rho$ at $\rho=a$ can be written as
\begin{eqnarray}\label{a4} 
{\cal J}_0(qa)=\nu_1{\cal J}_0(ka)+\nu_2{\cal N}_0(ka), \nonumber \\
q{\cal J}_1(qa)=\nu_1k{\cal J}_1(ka)+\nu_2k{\cal N}_1(ka).
\end{eqnarray}
Solving this system yields the following expression for $\varphi_p$
\begin{equation}\label{a5}
\tan\varphi_p={\cal J}_0(ka)\frac{q\frac{{\cal J}_1(qa)}{{\cal J}_0(qa)}-
k\frac{{\cal J}_1(ka)}{{\cal J}_0(ka)}}{q\frac{{\cal J}_1(qa)}{{\cal J}_0(qa)}
{\cal N}_0(ka)-k{\cal N}_1(ka)}.
\end{equation}
The case we are interested in is $ka\ll 1$. This allows to use the
small--argument asymptotics' for the functions depending on $ka$. One gets
\begin{equation}\label{a6}
\tan\varphi_p = \frac{\pi}{4}\left[\frac{2qa\frac{{\cal J}_1(qa)}{{\cal J}_0(qa)}-
k^2a^2}{1+qa\frac{{\cal J}_1(qa)}{{\cal J}_0(qa)}\ln\frac{2}{\gamma ka}}\right],
\end{equation} 
where $\gamma$ is the Euler constant. In the limit of a point--like impurity,
 considered by Gredeskul and Azbel\cite{GrAz}, we have 
$V_0<\hbar^2/ma^2$ (which is equivalent to $qa<1$), and Eq.(\ref{a6})
simplifies to
\begin{equation}\label{a7}
\tan\varphi_p = -\frac{\pi ma^2}{2\hbar^2}\left[\frac{V_0}
{1+\frac{ma^2}{\hbar^2}(V_0-E_p)\ln\frac{2}{\gamma ka}}\right].
\end{equation}
We  see that $V_0$ enters into the phase shift in the same ``renormalized''
form as in Eq. (\ref{30}) for $\tilde V_0^R$. If the renormalization is weak, 
we get $\varphi_p=-\pi ma^2V_0/2\hbar^2$, which leads to the standard result,
$V_0a^2/2l^2$, for the binding energy. As $V_0$ exceeds $\hbar^2/ma^2\ln
\frac{2}{\gamma ka}$ (but remains smaller than $\hbar^2/ma^2$) we get from
(\ref{a7})
\begin{equation}\label{a8}
\varphi_p=-\frac{\pi}{2\ln\frac{2}{\gamma ka}},
\end{equation}
so that $V_0$ disappears from the phase shift and, correspondingly, from the
binding energy. This is equivalent to the conclusion that $\tilde V_0^R$
renormalizes to $K^{-1}$. 

Most importantly, Eq.(\ref{a8}) remains valid when $V_0$ gets {\em much larger}
than $\hbar^2/ma^2$, and the approach, adopted in Sec. \ref{sec:H} ,
 is not applicable
any more. Indeed, as $qa$ increases, the combination 
$qa\frac{{\cal J}_1(qa)}{{\cal J}_0(qa)}$ becomes either oscillating
(for $V_0<0$) or monotonously increasing (for $V_0>0$)
function with a typical magnitude much larger than unity. Then the
result (\ref{a8}) immediately follows from (\ref{a6}).

\begin{figure}
\caption{Dimensionless functions $\Theta$ (solid curve) and $\Sigma$
(long-dashed curve) are shown versus the dimensionless energy $\varepsilon$
for the impurity position right at the center of the saddle point.
The dotted curve represents the ratio $|\Sigma|e^{-\frac{\pi |\varepsilon|}{2}}
/\Theta$.}
\label{fig1}
\end{figure}
\begin{figure}
\caption{The transmission coefficient as a function of energy 
for the case  $X_0=Y_0=0$ is shown for different values of the renormalized
impurity potential $\tilde V_0^R$: $\tilde V_0^R = -5$ (solid curve),
$\tilde V_0^R = -3$ (long-dashed curve), and $\tilde V_0^R = -1$ (dashed-dotted curve).}
\label{fig2}
\end{figure}
\begin{figure}
\caption{The transmission coefficient at zero energy is shown as a function
of the dimensionless impurity potential
 $v= \tilde V_0^R |\Gamma(1/4)|^2/4(2 \pi)^{1/2}$.}
\label{fig3}
\end{figure}

\begin{references}
\bibitem{Gloz1} I. Glozman, C. E. Johnson, and H. W. Jiang, Phys. Rev. Lett.
{\bf 74}, 594 (1995).
\bibitem{Gloz2} I. Glozman, C. E. Johnson and H. W. Jiang, 
Phys. Rev. B {\bf 52}, 14348 (1995).
\bibitem{Gloz3} I. Glozman, C. E. Johnson and H. W. Jiang, in 
{\em Proceedings of the 6th International Conference on Hopping and Related Phenomena}, ed. by O. Millo and Z. Ovadyahu, Jerusalem (1995).
\bibitem{Kra} S. V. Kravchenko, W. Mason, J. E. Furneaux, and V.M. Pudalov,
Phys. Rev. Lett. {\bf 75}, 910 (1995).
\bibitem{Fur} J. E. Furneaux, S. V. Kravchenko, W. E. Mason, G. E. Bowker,
and V. M. Pudalov, Phys. Rev. B {\bf 51}, 17227 (1995).
\bibitem{SR} T. V. Shahbazyan and M. E. Raikh, Phys. Rev. Lett. {\bf 75},
304 (1995).
\bibitem{Liu} D. Z. Liu, X. C. Xie, and Q. Niu, preprint (cond-mat/9504010).
\bibitem{Kag} V. Kagalovsky, B. Horovitz,  and Y. Avishai,  in 
{\em Proceedings of the 6th International Conference on Hopping and Related Phenomena}, ed. by O. Millo and Z. Ovadyahu, Jerusalem (1995). 
\bibitem{Khm} D. E. Khmelnitskii, Phys. Lett. {\bf 105A}, 182 (1984).
\bibitem{L} R. B. Laughlin, Phys. Rev. Lett. {\bf 52}, 2304 (1984).
\bibitem{LLP} H. Levine, S. B. Libby, and A. M. M. Pruisken,
Phys. Rev. Lett. {\bf 51}, 1915 (1983).
\bibitem{KLZ} S. Kivelson, D.-H. Lee, and S.-C. Zhang, Phys. Rev. B
{\bf 46}, 2223 (1992).
\bibitem{J} H. W. Jiang, C. E. Johnson, K. L. Wang, and S. T. Hannahs,
Phys. Rev. Lett. {\bf 71}, 1439 (1993). 
\bibitem{JJ} C. E. Johnson and H. W. Jiang, Phys. Rev. B {\bf 48}, 2823 (1993).
\bibitem{W}T. Wang, K. P. Clark, G. F. Spencer, A. M. Mack, and W. P. Kirk,
Phys. Rev. Lett. {\bf 72}, 709 (1994).
\bibitem{H} R. J. F. Hughes, J. T. Nicholls, J. E. F. Frost, E. H. Linfield, 
M. Pepper, C. J. B. Ford, D. A. Ritchie, G. A. C. Jones, E. Kogan, and
M. Kaveh, J. Phys. Condens. Matter {\bf 6}, 4763 (1994).
\bibitem{Kun} K. Yang and R. N. Bhatt, preprint (cond-mat/9510066).
\bibitem{An} T. Ando, J. Phys. Soc. Jpn. {\bf 53}, 3126 (1984).
\bibitem{CC} J. T. Chalker and P. D. Coddington, J. Phys. C {\bf 21},
2665 (1988).
\bibitem{fer} H. Fertig and B. I. Halperin, Phys. Rev. B {\bf 36}, 7969 (1987).
\bibitem{A.E} {\it Higher Transcendental Functions}, 
edited by A. Erd\'{e}lyi, (McGraw-Hill, New York, 1953), Vol. 2.
\bibitem{A} T. Ando, J. Phys. Soc. Jpn. {\bf 36}, 1521 (1974).
\bibitem{B} E. M. Baskin, L. N. Magarill, and M. V. Entin, Zh. Eksp. Teor.
Fiz. {\bf 75}, 723 (1978) [Sov. Phys. JETP {\bf 48}, 365 (1978)].
\bibitem{GM} V. A. Geiler and V. A. Margulis, Teor. Mat. Fiz. {\bf 58},
461 (1984) [Theor. Math. Phys. {\bf 58}, 302 (1984)].
\bibitem{Gr} S. A. Gredeskul, Y. Avishai, and M. Ya. Azbel', Europhys. Lett.
 {\bf 21}, 489 (1993).
\bibitem{Az} M. Ya. Azbel' and B. I. Halperin, Phys. Rev. B {\bf 52}, 14098
(1995).
\bibitem{GrAz} S. A. Gredeskul and M. Ya. Azbel' Phys. Rev. B {\bf 49}, 2323
(1994). 
\end{references}
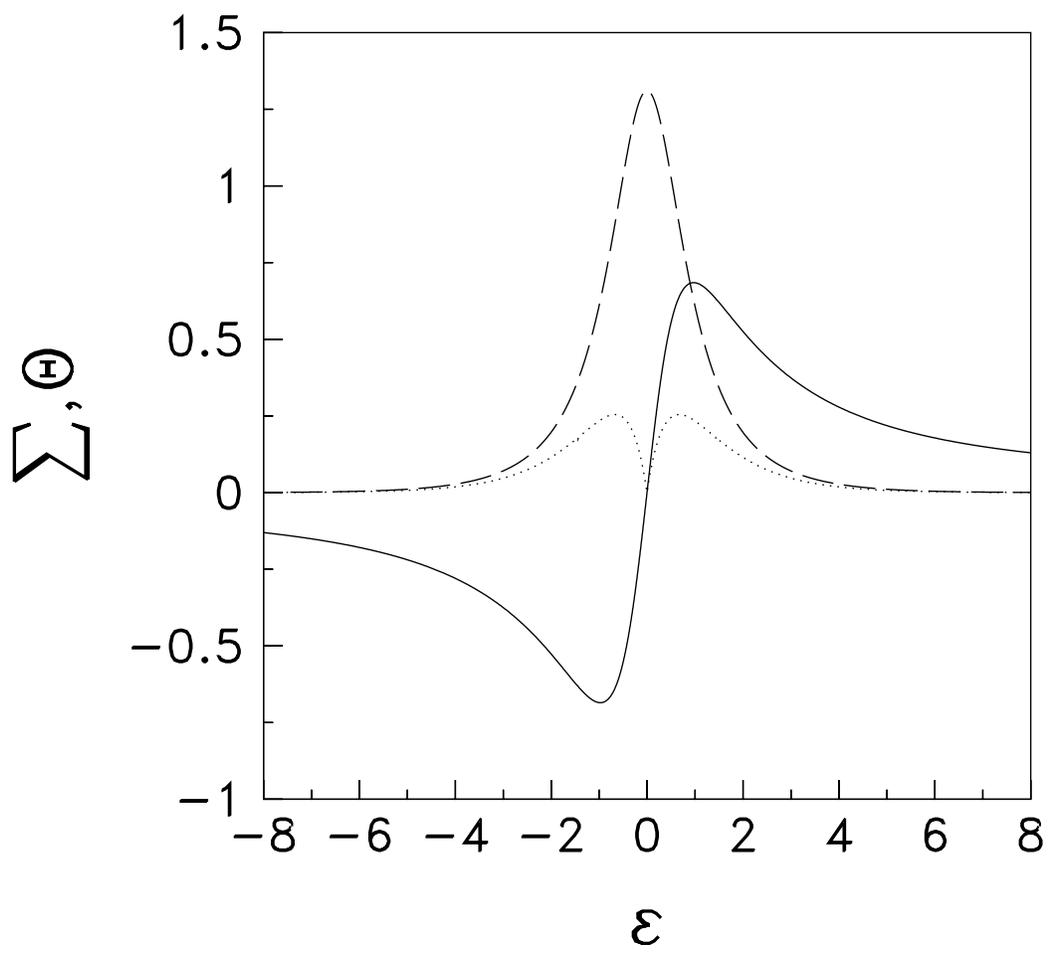
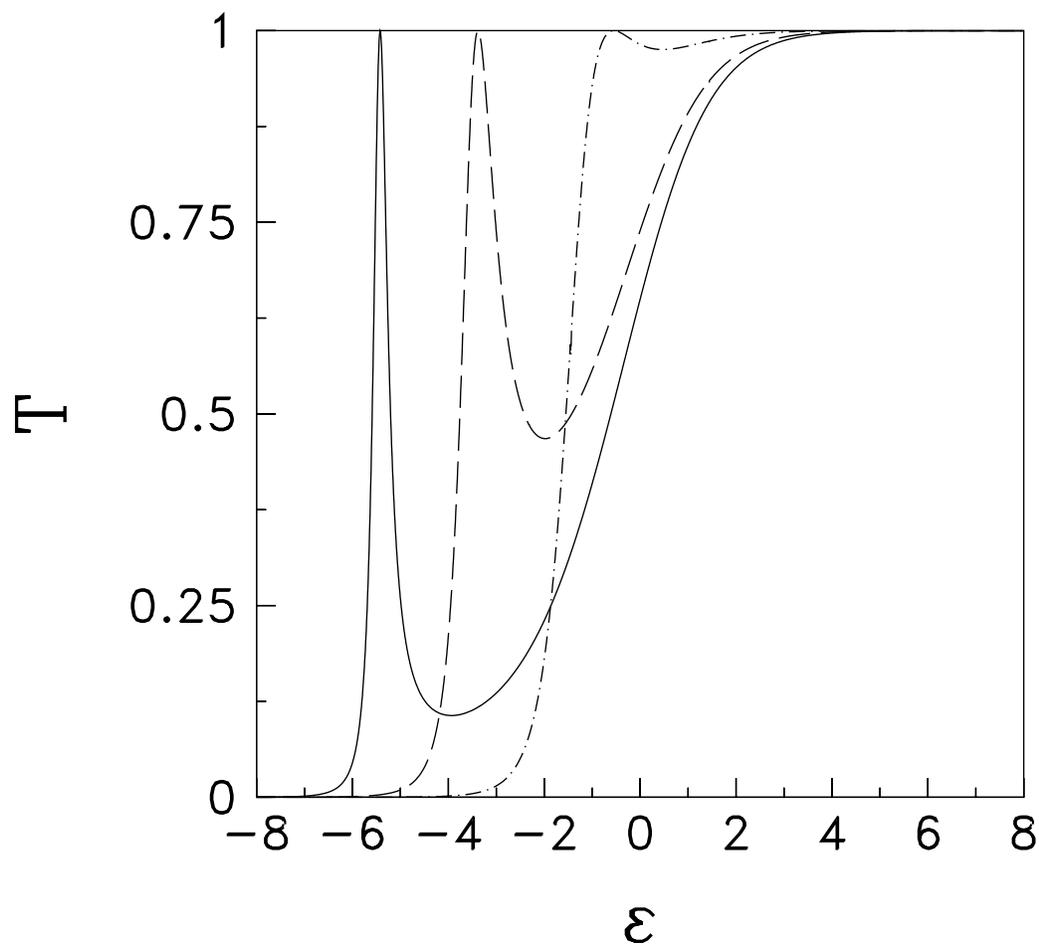
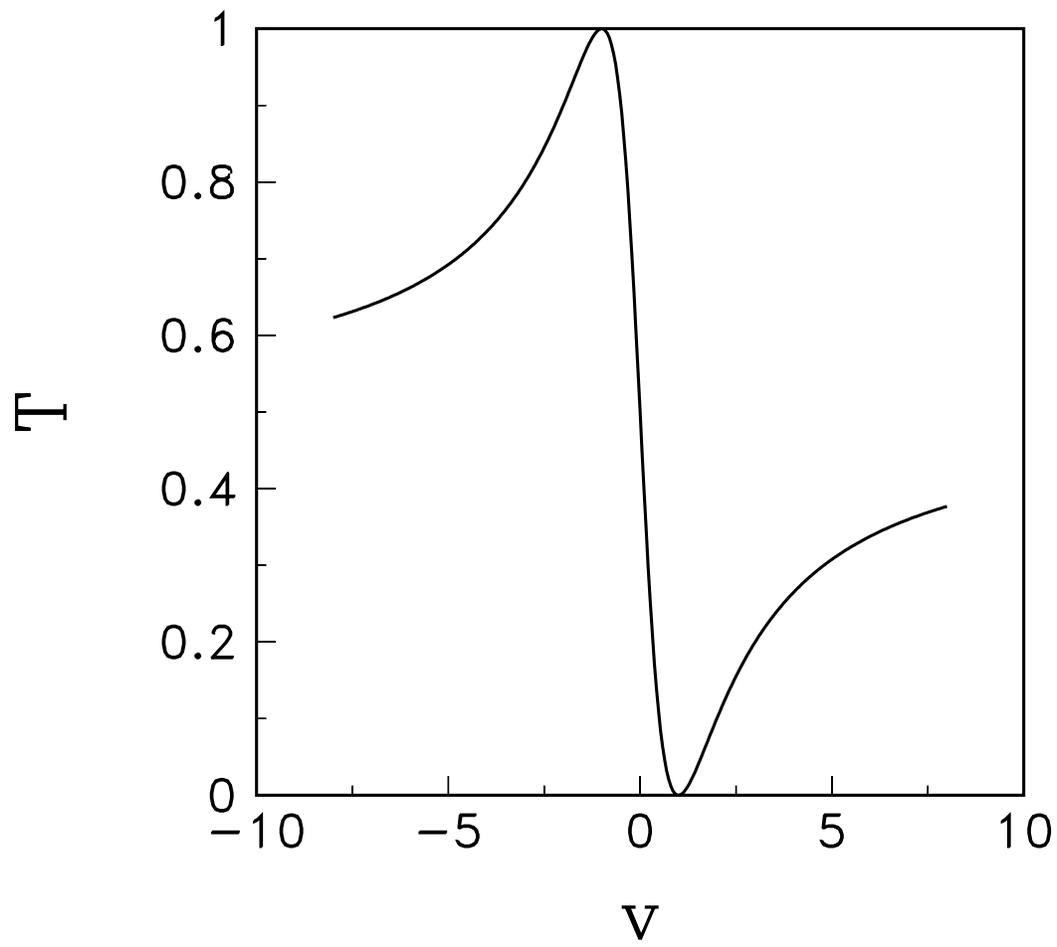
\end{document}